# Co-occurrence Matrices and their Applications in Information Science: Extending ACA to the Web Environment

Loet Leydesdorff [1] and Liwen Vaughan [2]

**Abstract**

Co-occurrence matrices, such as co-citation, co-word, and co-link matrices, have been used widely in the information sciences. However, confusion and controversy have hindered the proper statistical analysis of this data. The underlying problem, in our opinion, involved understanding the nature of various types of matrices. This paper discusses the difference between a symmetrical co-citation matrix and an asymmetrical citation matrix as well as the appropriate statistical techniques that can be applied to each of these matrices, respectively. Similarity measures (like the Pearson correlation coefficient or the cosine) should not be applied to the symmetrical co-citation matrix, but can be applied to the asymmetrical citation matrix to derive the proximity matrix. The argument is illustrated with examples. The study then extends the application of co-occurrence matrices to the Web environment where the nature of the available data and thus data collection methods are different from those of traditional databases such as the *Science Citation Index*. A set of data collected with the Google Scholar search engine is analyzed using both the traditional methods of multivariate analysis and the new visualization software Pajek that is based on social network analysis and graph theory.

[1] School of Economics (HEC), University of Lausanne, Switzerland & Amsterdam School of Communications Research (ASCoR), University of Amsterdam, Kloveniersburgwal 48, 1012 CX Amsterdam, The Netherlands. http://www.leydesdorff.net; email: loet@leydesdorff.net
[2] Faculty of Information & Media Studies, University of Western Ontario, London, Ontario, Canada N6A 5B7. http://www.fims.uwo.ca/whoswho/facultypage.htm?PeopleId=125; email: lvaughan@uwo.ca

# 1. Introduction

Co-occurrence matrices, such as co-citation, co-word, and co-link matrices, provide us with useful data for mapping and understanding the structures in the underlying document sets. Various types of analysis have been carried out on this data and a significant body of literature has been built up, making it an important area of information science (e.g., White & McCain, 1998). However, confusion persists about the nature of these matrices and the kinds of analysis that are appropriate. For example, the debate between Ahlgren, Jarneving, & Rousseau (2003, 2004a and b), White (2003, 2004), and Bensman (2004) on the use of the Pearson correlation coefficient or the cosine in the case of author cocitation analysis (ACA) shows some of these problems. In our opinion, co-occurrence matrices like the ones used in ACA are proximity data which do not require conversion before mapping. We shall argue that it is advisable to use, if possible, the asymmetrical matrices of documents versus attributes from which the co-occurrence matrices can be derived for mapping purposes.

This problem of how to process and understand co-occurrence matrices has entered a new dimension because co-occurrence matrices can be used extensively in Internet research. In this environment, one often can no longer retrieve the entire document set that is needed to construct the co-occurrence matrix, but one can construct these matrices directly, for example, by searching in a domain with Boolean ANDs. We will discuss the nature of the various matrices and the issues surrounding their analysis in the hope of clarifying some confusion and thus contributing to the further development of this area of information science. Our argument is methodological, but we shall illustrate the argument by using the example of an author co-citation analysis (ACA) in information science based on the ISI database which has been the subject of a previous debate in this journal (Ahlgren *et al*., 2003, 2004a, 2004b; White, 2003, 2004; Bensman, 2004; Leydesdorff, 2005). In the next step, we extend the data collection and analysis to the Web environment by using the Google Scholar search engine for the same set of scholars.

**2. Symmetrical Co-citation Matrix vs. Asymmetrical Citation Matrix**

*2.1    The symmetrical co-citation matrix*

Small (1973) pioneered co-citation analysis (cf. Marshakova, 1973). He constructed co-citation matrices in the form shown in Figure 1. The number in each cell of the matrix is the number of times two papers are co-cited. For example, Paper 1 and Paper 2 are co-cited 10 times while Paper 1 and Paper 3 are co-cited 20 times. At that time (early 1970s) Small had to use the ISI data as lists instead of matrices because of computational constraints. Using single linkage clustering Small could extract co-citation maps from this data without generating the matrices (Leydesdorff, 1987).

|         | Paper 1 | Paper 2 | Paper 3 | Paper 4 |
|---------|---------|---------|---------|---------|
| Paper 1 |         | 10      | 20      | 25      |
| Paper 2 | 10      |         | 30      | 15      |
| Paper 3 | 20      | 30      |         | 12      |
| Paper 4 | 25      | 15      | 12      |         |

**Figure 1:** Co-citation matrix (symmetrical matrix)

White & Griffith (1981) extended the co-citation analysis concept to author co-citation analysis (ACA), making significant contributions to the development of the field. They used first authors, rather than papers, as the units for analysis; as against co-citation analysis, the cited *authors*, not the cited *documents*, were their units of analysis. So their matrix is essentially the same as shown in Figure 1 except that Paper 1, Paper 2 etc. are replaced with Author 1, Author 2 etc. Both Small (1973) and White & Griffith (1981) used multidimensional scaling (MDS) and cluster analysis to analyze their data. White & McCain (1998) also used factor analysis. The difference is that Small normalized the original data (the raw co-citation data collected from the ISI databases) with the Jaccard Index while White & Griffith used the Pearson correlation coefficient for this purpose. Small & Sweeny (1985) began to use the cosine as an alternative similarity measure (Salton & McGill, 1983).

A matrix in the form of Figure 1 is a proximity matrix. As Kruskal (1978, p. 7) formulated: "A proximity is a number which indicates how similar or how different two objects are, or are perceived to be, or any measure of this kind." Proximity matrices can be either similarity or dissimilarity matrices (Cox & Cox, 2001, p. 9). Co-citation or co-author matrices are similarity (not dissimilarity) matrices. The higher the number in the cell, the more similar two papers (or two authors) appear to be. A proximity matrix can be input into multidimensional scaling software directly to generate a map which shows the relative positions of the papers or authors. The mapping principle is that the higher the proximity (the more similar two units are), the closer the two papers or authors will be located in the map.

## 2.2 The asymmetrical citation matrix

An alternative way of using citation data is to construct a matrix in the form shown in Figure 2. We will show an example of using this matrix for author co-citation analysis later. In this matrix, the rows are the citing papers and the columns represent cited papers. So Paper A is cited by Paper 1, 4, and 5 while C is cited by Paper 2 and 3.

|  | Cited Paper A | Cited Paper B | Cited Paper C | Cited Paper D |
|---|---|---|---|---|
| Citing Paper 1 | 1 | 1 | 0 | 1 |
| Citing Paper 2 | 0 | 0 | 1 | 1 |
| Citing Paper 3 | 0 | 0 | 1 | 1 |
| Citing Paper 4 | 1 | 1 | 0 | 0 |
| Citing Paper 5 | 1 | 1 | 0 | 1 |

**Figure 2:** Citation Matrix (asymmetrical matrix)

This matrix is VERY different from that shown in Figure 1. The matrix in Figure 1 is a symmetrical matrix in that (1) rows and columns are the same objects; (2) the number of rows is the same as the number of columns; and (3) data in the matrix is symmetrical about the diagonal, so that only half of the matrix is enough to contain all the data. Obviously, the matrix in Figure 2 does not have any of these three features, and this matrix is asymmetrical. Furthermore, data in the Figure 2 matrix are NOT proximity measures, so this matrix cannot be input directly into MDS. However, one can convert this attribute matrix into a proximity matrix. "A very common way to get proximities from data that are not proximities and hence inappropriate for MDS in their original form is to compute some measure of profile similarity or dissimilarity between rows (or columns) of a table. [...] The most common ways to derive a profile proximity measure are to compute correlations between variables or squared (Euclidean) distances between the stimuli" (Kruskal, 1978, p. 10). The Euclidean distance matrix can be considered as a dissimilarity matrix, while the Pearson correlation matrix can be considered as a similarity matrix. However, Ahlgren *et al.*'s (2003) argued that Pearson's correlation coefficient is formally not a similarity measure, but a measure of linear dependence. (See the discussion on similarity vs. dissimilarity matrices in the next section).

We focus here on the Pearson correlation coefficient, but a similar reasoning could be applied to the cosine as a similarity measure, or to Euclidean distances as dissimilarity measures (Ahlgren *et al.*, 2003, at p. 551). The problem of the potentially negative values of Pearson's $r$ as a proximity measure can be overcome by a linear transformation of $(r + 1) / 2$ which will result in values between 0 and 1. By applying the Pearson correlation to Figure 2 data (column pair wise correlation) and then using the conversion of $(r + 1) / 2$, one obtains the proximity matrix shown in Figure 3. This is a proximity matrix which has all the three features of the symmetry that Figure 1 has. Looking at data in Figure 2, we see that Paper A and Paper B are cited similarly by this set of papers (Paper 1, 4, and 5). The coefficient of 1 in Figure 3 reflects this fact. In contrast, Paper A and Paper C are cited completely dissimilarly in this set: they have the opposite citing papers, as shown by the coefficient of 0 in Figure 3.

|         | Paper A | Paper B | Paper C | Paper D |
|---------|---------|---------|---------|---------|
| Paper A | 1       | 1       | 0       | 0.295   |
| Paper B | 1       | 1       | 0       | 0.295   |
| Paper C | 0       | 0       | 1       | 0.705   |
| Paper D | 0.295   | 0.295   | 0.705   | 1       |

**Figure 3:** Proximity matrix derived from data in Figure 2

In the case of the asymmetrical matrix (Figure 2), the cited papers are considered as attributes of the citing papers because they are contained in the reference lists of the latter. Paper A shares two out of three of its citing papers with Paper D so their coefficient is a number between 0 and 1, i.e. 0.295.

In summary, the Pearson correlation coefficient can be used in co-citation analysis when it is applied to an asymmetrical citation matrix. However, applying Pearson's *r* to a symmetrical proximity matrix is problematic. White (2003, p. 1251) noted that on the first page of Davison's (1983) textbook on multidimensional scaling, the correlation coefficient is mentioned as one of the two basic proximity measures in MDS. However, in this book, Pearson correlation coefficients were always used to construct proximity matrices from data that were not already proximity measures. A co-citation matrix is a proximity matrix, so there is no need to apply a similarity measure to construct a proximity matrix. On the contrary, doing so may distort the data, as we will now show with an empirical example.

## 2.3 An example

The data for our example (see Table 1) were copied from SPSS (1993). This table provides flying mileages among ten American Cities. The map from which these distances are generated is two-dimensional and thus one can evaluate straightforwardly the quality of the reconstruction of the geographical map using this data.

**Table 1.** Flying mileages between 10 American Cities

|  | Atlanta | Chicago | Denver | Houston | Los Angeles | Miami | New York | San Francisco | Seattle | Washington DC |
|---|---|---|---|---|---|---|---|---|---|---|
| Atlanta | 0 | . | . | . | . | . | . | . | . | . |
| Chicago | 587 | 0 | . | . | . | . | . | . | . | . |
| Denver | 1212 | 920 | 0 | . | . | . | . | . | . | . |
| Houston | 701 | 940 | 879 | 0 | . | . | . | . | . | . |
| Los Angeles | 1936 | 1745 | 831 | 1374 | 0 | . | . | . | . | . |
| Miami | 604 | 1188 | 1726 | 968 | 2339 | 0 | . | . | . | . |
| New York | 748 | 713 | 1631 | 1420 | 2451 | 1092 | 0 | . | . | . |
| San Francisco | 2139 | 1858 | 949 | 1645 | 347 | 2594 | 2571 | 0 | . | . |
| Seattle | 2182 | 1737 | 1021 | 1891 | 959 | 2734 | 2408 | 678 | 0 | . |
| Washington DC | 543 | 597 | 1494 | 1220 | 2300 | 923 | 205 | 2442 | 2329 | 0 |

Obviously, this is a symmetrical proximity matrix. The data measures *dissimilarity*, as the larger the numbers, the further apart the cities are, i.e. the more "dissimilar" they are in location. By inputting this matrix into SPSS and choosing PROXSCAL as an option of MDS, we obtain Figure 3, which is an almost perfect mapping of the relative positions of these cities (the positions are relative and the map is reversed in terms of west and east. However, because of this relativity of the positions the results of MDS can be rotated freely for the interpretation).

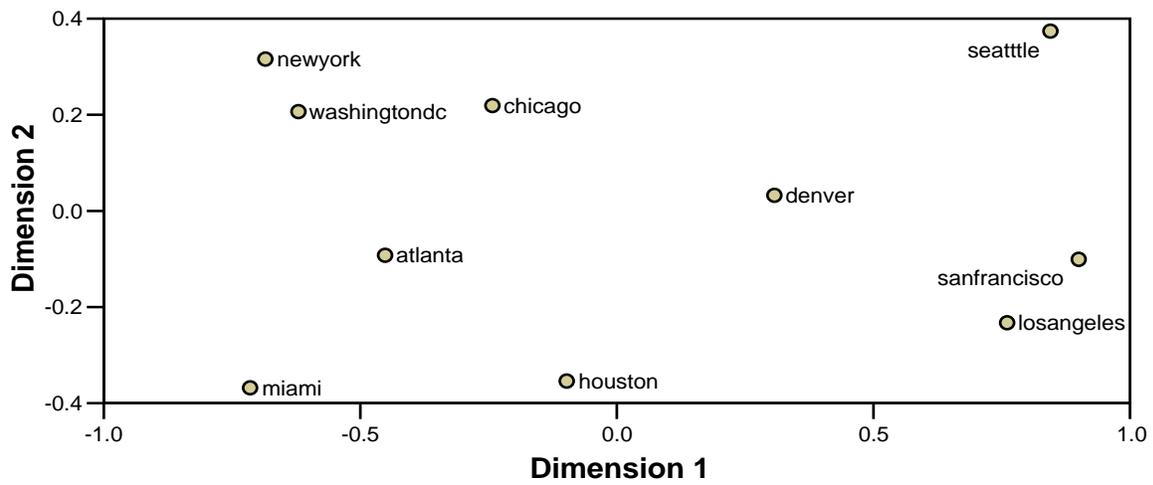

**Figure 4:** MDS mapping (PROXSCAL) of ten American cities using the original distance matrix (normalized raw stress = 0.0001)

After applying Pearson's *r* to the data of Table 1 and then map this new matrix with MDS, we obtain a distorted map of the ten cities and the normalized raw stress of this picture is very high (0.11341).

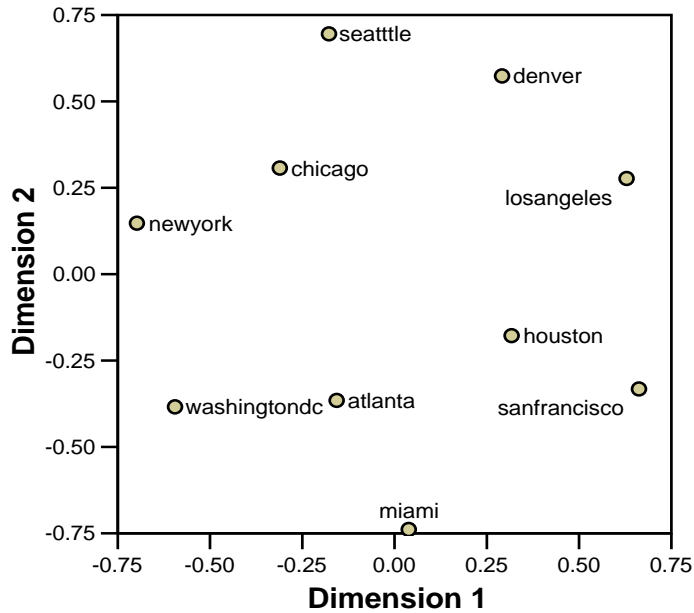

**Figure 5:** MDS mapping of ten American cities using the Pearson correlation matrix of the distances (normalized raw stress = 0.11341)

Apparently, Figure 5 does not improve on Figure 4 (the stress has become very high). By using the Pearson correlations instead of the distances, the representation is distorted. For example, Los Angeles is positioned closer to Seattle than San Francisco while New York is closer to Chicago than to Washington, D.C. The Pearson correlation normalizes the data with reference to the mean, and the pattern of co-occurrences as variables, as indicated by the Pearson correlation, is in some cases different from the proximities in the network.

Unlike this geographical data—which is two-dimensional and therefore can be mapped unambiguously—the intellectual structure as measured, for example, by using co-authorship or co-citation data is usually multi-dimensional. Multi-dimensional scaling (or factor analysis) searches for a projection of the n-dimensional data in a space with lower dimensionality. MDS uses the stress measure as an indicator for the fit, but this can only be considered as a heuristic.

Eventually, the analyst also has to appreciate the representation of the represented structure on qualitative grounds. In other words, the multi-dimensional representation of intellectual structure in terms of co-authorship data can be very good, while this representation cannot easily be projected in a two- or three-dimensional visualization. Factor analysis allows us to study the quality of data reduction in more dimensions in precise numbers (algorithmically) and may thus be helpful in understanding the quality of the geometrical visualization as a projection.

**3. Similarity vs. dissimilarity measures**

As stated above, there are two kinds of proximity measures: similarity or dissimilarity. Obviously the two are opposite, so they should be treated differently in MDS. In recent versions of SPSS, there are two options for MDS: ALSCAL and PROXSCAL. ALSCAL assumes that the input is a dissimilarity matrix, while PROXSCAL allows one to specify whether the proximities are similarity or dissimilarity measures. There is no doubt that co-citation is a similarity measure (the more co-citations two papers or two authors have, the more similar they are), so the similarity option of PROXSCAL should be used. If one reverses the two types of similarity measures, the mapping results will be wrong. For example, the America city mileage data in Table 1 provides a dissimilarity measure and if we specify it as a similarity measure in MDS, the result is a very distorted map (the resulting map is omitted here due to space limitations).

In early versions of SPSS, only the ALSCAL option was available (the dissimilarity measure only). In this case, a co-citation matrix should be converted into a dissimilarity matrix before it is input into SPSS. Kruskal & Wish (1978, p. 77) clearly state that "If the proximities are similarities, they must be 'turned upside down' into dissimilarities, for example by forming dissimilarity = (constant – similarity) where the value of the constant is judiciously chosen." If the similarity measure is between 0 and 1 (e.g. the above example of using Pearson's *r* to obtain the proximity matrix of Figure 3), then the constant can be 1, i.e. dissimilarity = (1 – similarity). One of us conducted extensive testing of the formulae and found that the mapping results from using dissimilarity measures after the correct conversion from similarity to dissimilarity, and from using the similarity measures directly, are always the same.

A widely used form of MDS is on asymmetrical attribute matrices as in Figure 2 above. MDS is then primarily a visualization technique within a class of multivariate instruments like factor analysis, cluster analysis, etc. In this case, the data is analyzed as dissimilar variables, and thus both ALSCAL and PROXSCAL can be used. Euclidean distances are the default measure of dissimilarity. For input data that are not proximity measures, PROXSCAL can construct the proximity matrix. Because we study both types of matrices in the various sections below, we will use PROXSCAL throughout this study. Note that a visualization such as MDS remains a representation of the data in two or three dimensions, while factor analysis, for example, adds the possibility of rotating the data in order to obtain a higher-dimensional and quantitative understanding of the structures underlying these geometrical representations (Schiffman *et al.*, 1981).

## 4. An example of Author Co-citation Analysis (ACA)

Let us return to the example of an author co-citation analysis that was discussed previously in several contributions to this journal (Ahlgren *et al.*, 2003; White, 2003; Bensman, 2004; Leydesdorff, 2005) and discuss in considerable detail the consequences of not using the symmetrical co-occurrence matrix but rather the asymmetrical matrix of documents versus references.

Ahlgren *et al.* (2003: 554) downloaded from the *Web of Science* 430 bibliographic descriptions of articles published in *Scientometrics* and 483 such descriptions published in the *Journal of the American Society for Information Science and Technology (JASIST)* in the period 1996-2000. From the 913 bibliographic references in these articles they composed a co-occurrence matrix for twelve authors in the field of information retrieval and 12 authors doing bibliometric-scientometric research. They provide both the co-occurrence matrix and the Pearson correlation table in their paper (at pp. 555 and 556, respectively).

We repeated the analysis in order to obtain the original (asymmetrical) data matrix. Using precisely the same searches we found 469 articles in *Scientometrics* and 494 in *JASIST* on 18 November 2004. The somewhat higher numbers are consistent with the practice of the ISI to reallocate papers sometimes at a later date to a previous year. Thus, we disregarded these differences.

*4.1 Descriptive statistics*

Of the (469 + 494 =) 963 documents thus retrieved, 902 contain 21,813 references. 279 records contain at least one co-citation to two or more authors of the list of 24 authors under study. There are no citing records which contain a reference to only a single author in this set of 279 citing documents. Thus, this can with good reason be considered as a set of highly co-cited authors. Figure 6 shows that one citing paper even co-cited ten of the authors included in the analysis.

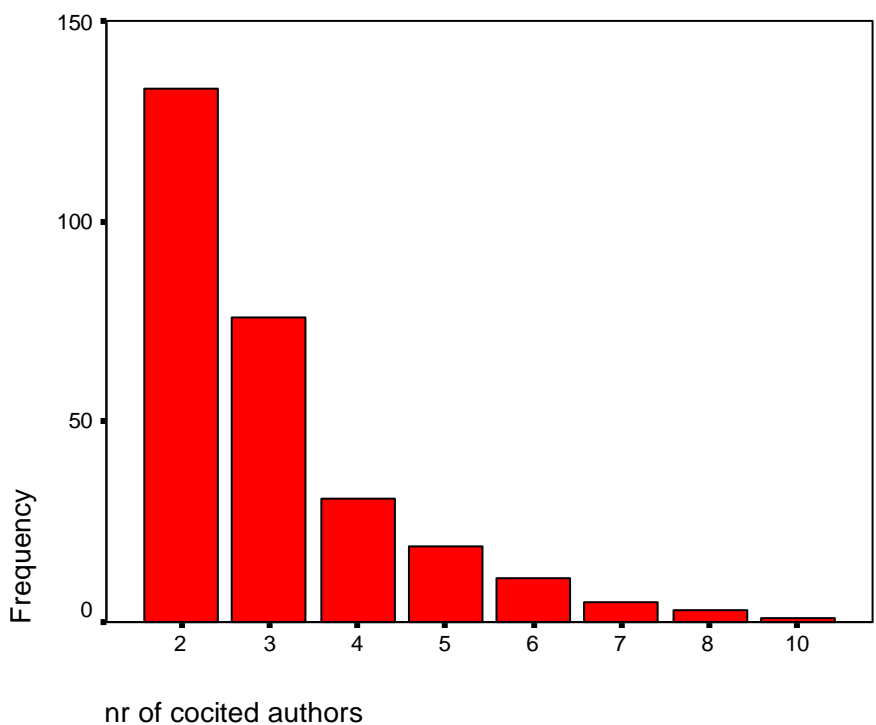

**Figure 6:** Distribution of 279 co-citations in terms of the number of authors co-cited in a single citing document

Figure 7 exhibits the total citations of these authors within the set of citing documents. Note that the scientometrics authors have on average a citation rate of 44.6 (± 14.8), while the information retrieval researchers have a lower average of 26.1 (± 6.5). Citation rates are field-specific, indeed.

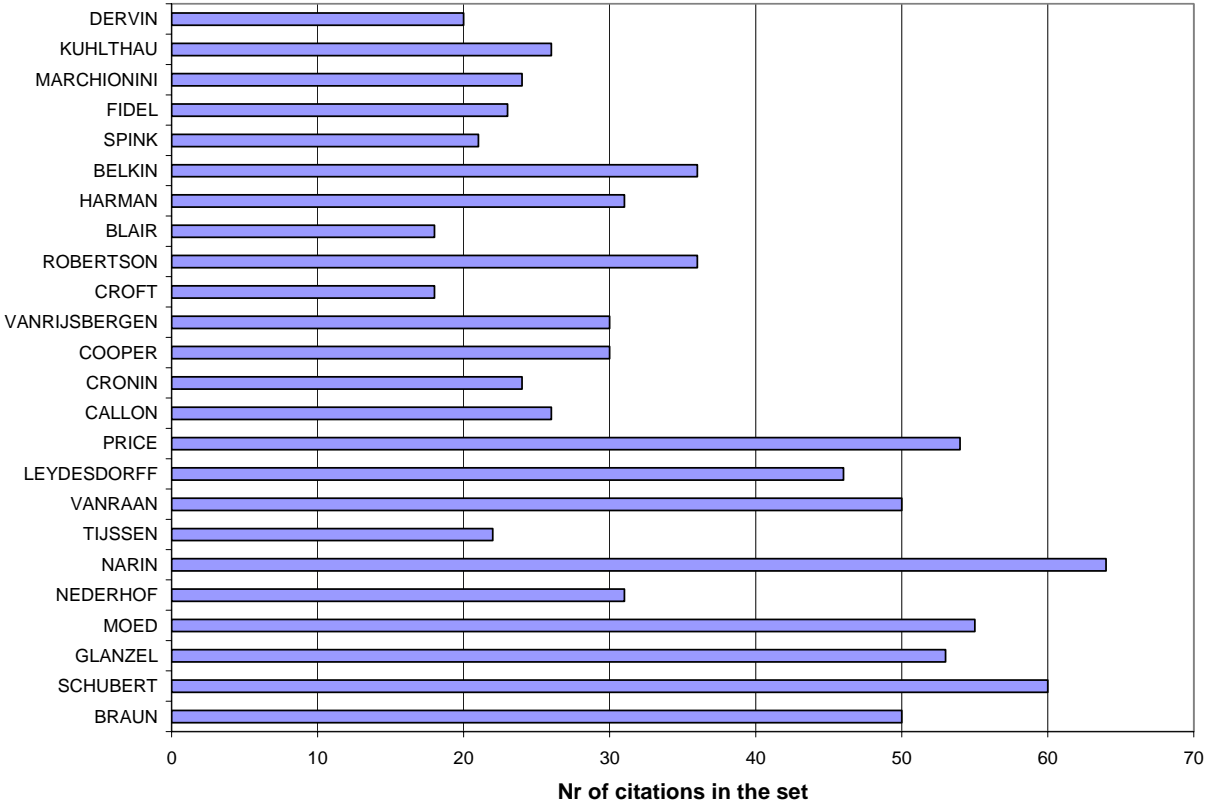

**Figure 7.** Number of times each of the 24 authors is cited in the 279 citing documents

Let us now move on from these descriptive statistics to an analysis of the data.

*4.2    Data analysis of the asymmetrical matrix*

The data can be imported into SPSS and the asymmetrical matrix can then be subjected to various forms of multivariate analysis. For example, one can ask for a Pearson correlation matrix. Table 2 provides this matrix for our 24 authors. These Pearson correlations are very different from the ones provided for this set of authors by Ahlgren *et al*. (2003, at p. 556), since they applied Pearson's *r* to the symmetrical co-citation matrix. For example, the correlation coefficient between the co-citation pattern of Van Raan and Schubert in the latters' Table 9 is 0.74, while we found a negative correlation between their citation patterns ($r = -0.131$; $p < 0.05$). The Pearson correlations derived from the symmetrical co-citation matrix are all high and significant *because* this matrix is symmetrical, so all values and relations occur twice.

**Table 2** Pearson correlations among the 24 cited authors on the basis of 279 citing documents

**Correlations**

| | | BRAUN | SCHUBERT | GLANZEL | MOED | NEDERHOF | NARIN | TIJSSEN | VANRAAN | LEYDESDORFF | PRICE | CALLON | CRONIN | COOPER | VANRIJSBERGEN | CROFT | ROBERTSON | BLAIR | HARMAN | BELKIN | SPINK | FIDEL | MARCHIONINI | KUHLTHAU | DERVIN |
|---|---|---|---|---|---|---|---|---|---|---|---|---|---|---|---|---|---|---|---|---|---|---|---|---|---|
| BRAUN | Pearson Correlation | 1 | .415** | .226** | .238** | .102 | .034 | .037 | .001 | -.031 | -.063 | -.085 | -.143* | -.162** | -.162** | -.123* | -.180** | -.123* | -.165** | -.180** | -.133* | -.140* | -.143* | -.150* | -.130* |
| | Sig. (2-tailed) | . | .000 | .000 | .000 | .088 | .572 | .542 | .987 | .602 | .292 | .154 | .017 | .007 | .007 | .041 | .003 | .041 | .006 | .003 | .026 | .019 | .017 | .012 | .030 |
| | N | 279 | 279 | 279 | 279 | 279 | 279 | 279 | 279 | 279 | 279 | 279 | 279 | 279 | 279 | 279 | 279 | 279 | 279 | 279 | 279 | 279 | 279 | 279 | 279 |
| SCHUBERT | Pearson Correlation | .415** | 1 | .369** | .135* | .093 | .088 | .009 | -.131* | -.115 | .009 | -.108 | -.129* | -.182** | -.182** | -.137* | -.201** | -.137* | -.185** | -.201** | -.149* | -.157** | -.161** | -.168** | -.145* |
| | Sig. (2-tailed) | .000 | . | .000 | .024 | .123 | .143 | .885 | .029 | .055 | .887 | .072 | .031 | .002 | .002 | .022 | .001 | .022 | .002 | .001 | .013 | .009 | .007 | .005 | .015 |
| | N | 279 | 279 | 279 | 279 | 279 | 279 | 279 | 279 | 279 | 279 | 279 | 279 | 279 | 279 | 279 | 279 | 279 | 279 | 279 | 279 | 279 | 279 | 279 | 279 |
| GLANZEL | Pearson Correlation | .226** | .369** | 1 | .128* | .120* | .170** | .163** | .107 | .006 | .017 | .002 | -.051 | -.168** | -.168** | -.127* | -.186** | -.127* | -.171** | -.186** | -.138* | -.145* | -.149* | -.155** | -.135* |
| | Sig. (2-tailed) | .000 | .000 | . | .033 | .046 | .004 | .006 | .074 | .915 | .775 | .975 | .398 | .005 | .005 | .034 | .002 | .034 | .004 | .002 | .021 | .015 | .013 | .009 | .025 |
| | N | 279 | 279 | 279 | 279 | 279 | 279 | 279 | 279 | 279 | 279 | 279 | 279 | 279 | 279 | 279 | 279 | 279 | 279 | 279 | 279 | 279 | 279 | 279 | 279 |
| MOED | Pearson Correlation | .238** | .135* | .128* | 1 | .169** | .158** | .022 | .191** | .095 | .031 | .058 | -.023 | -.172** | -.172** | -.130* | -.191** | -.130* | -.175** | -.191** | -.141* | -.149* | -.152* | -.159** | -.138* |
| | Sig. (2-tailed) | .000 | .024 | .033 | . | .005 | .008 | .712 | .001 | .112 | .607 | .334 | .696 | .004 | .004 | .030 | .001 | .030 | .003 | .001 | .018 | .013 | .011 | .008 | .021 |
| | N | 279 | 279 | 279 | 279 | 279 | 279 | 279 | 279 | 279 | 279 | 279 | 279 | 279 | 279 | 279 | 279 | 279 | 279 | 279 | 279 | 279 | 279 | 279 | 279 |
| NEDERHOF | Pearson Correlation | .102 | .093 | .120* | .169** | 1 | .133* | .235** | .162** | .058 | -.058 | .044 | -.027 | -.123* | -.123* | -.093 | -.136* | -.093 | -.125* | -.136* | -.101 | -.106 | -.108 | -.113 | -.098 |
| | Sig. (2-tailed) | .088 | .123 | .046 | .005 | . | .027 | .000 | .007 | .334 | .337 | .468 | .652 | .041 | .041 | .122 | .023 | .122 | .037 | .023 | .093 | .077 | .070 | .059 | .101 |
| | N | 279 | 279 | 279 | 279 | 279 | 279 | 279 | 279 | 279 | 279 | 279 | 279 | 279 | 279 | 279 | 279 | 279 | 279 | 279 | 279 | 279 | 279 | 279 | 279 |
| NARIN | Pearson Correlation | .034 | .088 | .170** | .158** | .133* | 1 | .188** | .190** | .240** | .164** | .148* | .106 | -.162** | -.189** | -.109 | -.185** | -.143* | -.193** | -.185** | -.123* | -.164** | -.167** | -.175** | -.152* |
| | Sig. (2-tailed) | .572 | .143 | .004 | .008 | .027 | . | .002 | .001 | .000 | .006 | .014 | .076 | .007 | .001 | .070 | .002 | .017 | .001 | .002 | .040 | .006 | .005 | .003 | .011 |
| | N | 279 | 279 | 279 | 279 | 279 | 279 | 279 | 279 | 279 | 279 | 279 | 279 | 279 | 279 | 279 | 279 | 279 | 279 | 279 | 279 | 279 | 279 | 279 | 279 |
| TIJSSEN | Pearson Correlation | .037 | .009 | .163** | .022 | .235** | .188** | 1 | .314** | .228** | .025 | .181** | -.042 | -.102 | -.059 | .031 | -.073 | -.077 | -.103 | -.113 | -.033 | -.088 | -.090 | -.094 | -.081 |
| | Sig. (2-tailed) | .542 | .885 | .006 | .712 | .000 | .002 | . | .000 | .000 | .678 | .002 | .481 | .090 | .329 | .601 | .225 | .201 | .085 | .060 | .582 | .144 | .135 | .118 | .176 |
| | N | 279 | 279 | 279 | 279 | 279 | 279 | 279 | 279 | 279 | 279 | 279 | 279 | 279 | 279 | 279 | 279 | 279 | 279 | 279 | 279 | 279 | 279 | 279 | 279 |
| VANRAAN | Pearson Correlation | .001 | -.131* | .107 | .191** | .162** | .190** | .314** | 1 | .120* | .055 | .236** | .057 | -.102 | -.132* | -.047 | -.152* | -.123* | -.165** | -.180** | -.098 | -.140* | -.143* | -.150* | -.130* |
| | Sig. (2-tailed) | .987 | .029 | .074 | .001 | .007 | .001 | .000 | . | .046 | .361 | .000 | .346 | .089 | .027 | .438 | .011 | .041 | .006 | .003 | .103 | .019 | .017 | .012 | .030 |
| | N | 279 | 279 | 279 | 279 | 279 | 279 | 279 | 279 | 279 | 279 | 279 | 279 | 279 | 279 | 279 | 279 | 279 | 279 | 279 | 279 | 279 | 279 | 279 | 279 |
| LEYDESDORFF | Pearson Correlation | -.031 | -.115 | .006 | .095 | .058 | .240** | .228** | .120* | 1 | .198** | .323** | .208** | -.123* | -.154** | -.077 | -.142* | -.117 | -.157** | -.171** | -.054 | -.133* | -.136* | -.142* | -.123* |
| | Sig. (2-tailed) | .602 | .055 | .915 | .112 | .334 | .000 | .000 | .046 | . | .001 | .000 | .000 | .040 | .010 | .198 | .017 | .052 | .009 | .004 | .373 | .026 | .023 | .017 | .039 |
| | N | 279 | 279 | 279 | 279 | 279 | 279 | 279 | 279 | 279 | 279 | 279 | 279 | 279 | 279 | 279 | 279 | 279 | 279 | 279 | 279 | 279 | 279 | 279 | 279 |
| PRICE | Pearson Correlation | -.063 | .009 | .017 | .031 | -.058 | .164** | .025 | .055 | .198** | 1 | .155** | .141* | -.141* | -.141* | -.092 | -.161** | -.129* | -.173** | -.134* | -.140* | -.114 | -.150* | -.126* | -.066 |
| | Sig. (2-tailed) | .292 | .887 | .775 | .607 | .337 | .006 | .678 | .361 | .001 | . | .009 | .019 | .019 | .019 | .126 | .007 | .032 | .004 | .025 | .020 | .058 | .012 | .036 | .273 |
| | N | 279 | 279 | 279 | 279 | 279 | 279 | 279 | 279 | 279 | 279 | 279 | 279 | 279 | 279 | 279 | 279 | 279 | 279 | 279 | 279 | 279 | 279 | 279 | 279 |
| CALLON | Pearson Correlation | -.085 | -.108 | .002 | .058 | .044 | .148* | .181** | .236** | .323** | .155** | 1 | .078 | -.111 | -.111 | -.034 | -.087 | -.084 | -.113 | -.123* | -.045 | -.096 | -.098 | -.103 | -.089 |
| | Sig. (2-tailed) | .154 | .072 | .975 | .334 | .468 | .014 | .002 | .000 | .000 | .009 | . | .197 | .063 | .063 | .572 | .149 | .161 | .059 | .039 | .457 | .109 | .101 | .087 | .138 |
| | N | 279 | 279 | 279 | 279 | 279 | 279 | 279 | 279 | 279 | 279 | 279 | 279 | 279 | 279 | 279 | 279 | 279 | 279 | 279 | 279 | 279 | 279 | 279 | 279 |
| CRONIN | Pearson Correlation | -.143* | -.129* | -.051 | -.023 | -.027 | .106 | -.042 | .057 | .208** | .141* | .078 | 1 | -.065 | -.106 | -.081 | -.080 | -.081 | -.108 | -.118* | -.039 | -.092 | -.049 | -.054 | -.036 |
| | Sig. (2-tailed) | .017 | .031 | .398 | .696 | .652 | .076 | .481 | .346 | .000 | .019 | .197 | . | .278 | .076 | .180 | .183 | .180 | .070 | .049 | .516 | .125 | .419 | .366 | .553 |
| | N | 279 | 279 | 279 | 279 | 279 | 279 | 279 | 279 | 279 | 279 | 279 | 279 | 279 | 279 | 279 | 279 | 279 | 279 | 279 | 279 | 279 | 279 | 279 | 279 |
| COOPER | Pearson Correlation | -.162** | -.182** | -.168** | -.172** | -.123* | -.162** | -.102 | -.102 | -.123* | -.141* | -.111 | -.065 | 1 | .440** | .144* | .281** | .144* | .245** | .108 | -.011 | -.104 | -.024 | -.071 | -.052 |
| | Sig. (2-tailed) | .007 | .002 | .005 | .004 | .041 | .007 | .090 | .089 | .040 | .019 | .063 | .278 | . | .000 | .016 | .000 | .016 | .000 | .072 | .851 | .083 | .690 | .234 | .391 |
| | N | 279 | 279 | 279 | 279 | 279 | 279 | 279 | 279 | 279 | 279 | 279 | 279 | 279 | 279 | 279 | 279 | 279 | 279 | 279 | 279 | 279 | 279 | 279 | 279 |
| VANRIJSBERGEN | Pearson Correlation | -.162** | -.182** | -.168** | -.172** | -.123* | -.189** | -.059 | -.132* | -.154** | -.141* | -.111 | -.106 | .440** | 1 | .239** | .453** | .144* | .356** | .039 | .033 | -.062 | -.106 | -.071 | -.052 |
| | Sig. (2-tailed) | .007 | .002 | .005 | .004 | .041 | .001 | .329 | .027 | .010 | .019 | .063 | .076 | .000 | . | .000 | .000 | .016 | .000 | .517 | .588 | .302 | .076 | .234 | .391 |
| | N | 279 | 279 | 279 | 279 | 279 | 279 | 279 | 279 | 279 | 279 | 279 | 279 | 279 | 279 | 279 | 279 | 279 | 279 | 279 | 279 | 279 | 279 | 279 | 279 |
| CROFT | Pearson Correlation | -.123* | -.137* | -.127* | -.130* | -.093 | -.109 | .031 | -.047 | -.077 | -.092 | -.034 | -.081 | .144* | .239** | 1 | .291** | .287** | .232** | .247** | .257** | .027 | -.029 | .016 | .040 |
| | Sig. (2-tailed) | .041 | .022 | .034 | .030 | .122 | .070 | .601 | .438 | .198 | .126 | .572 | .180 | .016 | .000 | . | .000 | .000 | .000 | .000 | .000 | .649 | .635 | .788 | .504 |
| | N | 279 | 279 | 279 | 279 | 279 | 279 | 279 | 279 | 279 | 279 | 279 | 279 | 279 | 279 | 279 | 279 | 279 | 279 | 279 | 279 | 279 | 279 | 279 | 279 |
| ROBERTSON | Pearson Correlation | -.180** | -.201** | -.186** | -.191** | -.136* | -.185** | -.073 | -.152* | -.142* | -.161** | -.087 | -.080 | .281** | .453** | .291** | 1 | .204** | .306** | .235** | .295** | .196** | .111 | .024 | .059 |
| | Sig. (2-tailed) | .003 | .001 | .002 | .001 | .023 | .002 | .225 | .011 | .017 | .007 | .149 | .183 | .000 | .000 | .000 | . | .001 | .000 | .000 | .000 | .001 | .065 | .693 | .328 |
| | N | 279 | 279 | 279 | 279 | 279 | 279 | 279 | 279 | 279 | 279 | 279 | 279 | 279 | 279 | 279 | 279 | 279 | 279 | 279 | 279 | 279 | 279 | 279 | 279 |
| BLAIR | Pearson Correlation | -.123* | -.137* | -.127* | -.130* | -.093 | -.143* | -.077 | -.123* | -.117 | -.129* | -.084 | -.081 | .144* | .144* | .287** | .204** | 1 | .371** | .073 | .036 | .027 | .023 | -.084 | -.073 |
| | Sig. (2-tailed) | .041 | .022 | .034 | .030 | .122 | .017 | .201 | .041 | .052 | .032 | .161 | .180 | .016 | .016 | .000 | .001 | . | .000 | .224 | .553 | .649 | .696 | .161 | .224 |
| | N | 279 | 279 | 279 | 279 | 279 | 279 | 279 | 279 | 279 | 279 | 279 | 279 | 279 | 279 | 279 | 279 | 279 | 279 | 279 | 279 | 279 | 279 | 279 | 279 |
| HARMAN | Pearson Correlation | -.165** | -.185** | -.171** | -.175** | -.125* | -.193** | -.103 | -.165** | -.157** | -.173** | -.113 | -.108 | .245** | .356** | .232** | .306** | .371** | 1 | .170** | .115 | .101 | .014 | -.074 | -.054 |
| | Sig. (2-tailed) | .006 | .002 | .004 | .003 | .037 | .001 | .085 | .006 | .009 | .004 | .059 | .070 | .000 | .000 | .000 | .000 | .000 | . | .004 | .054 | .091 | .822 | .217 | .369 |
| | N | 279 | 279 | 279 | 279 | 279 | 279 | 279 | 279 | 279 | 279 | 279 | 279 | 279 | 279 | 279 | 279 | 279 | 279 | 279 | 279 | 279 | 279 | 279 | 279 |
| BELKIN | Pearson Correlation | -.180** | -.201** | -.186** | -.191** | -.136* | -.185** | -.113 | -.180** | -.171** | -.134* | -.123* | -.118* | .108 | .039 | .247** | .235** | .073 | .170** | 1 | .255** | .235** | .263** | .392** | .308** |
| | Sig. (2-tailed) | .003 | .001 | .002 | .001 | .023 | .002 | .060 | .003 | .004 | .025 | .039 | .049 | .072 | .517 | .000 | .000 | .224 | .004 | . | .000 | .000 | .000 | .000 | .000 |
| | N | 279 | 279 | 279 | 279 | 279 | 279 | 279 | 279 | 279 | 279 | 279 | 279 | 279 | 279 | 279 | 279 | 279 | 279 | 279 | 279 | 279 | 279 | 279 | 279 |
| SPINK | Pearson Correlation | -.133* | -.149* | -.138* | -.141* | -.101 | -.123* | -.033 | -.098 | -.054 | -.140* | -.045 | -.039 | -.011 | .033 | .257** | .295** | .036 | .115 | .255** | 1 | .458** | .252** | .142* | .131* |
| | Sig. (2-tailed) | .026 | .013 | .021 | .018 | .093 | .040 | .582 | .103 | .373 | .020 | .457 | .516 | .851 | .588 | .000 | .000 | .553 | .054 | .000 | . | .000 | .000 | .017 | .028 |
| | N | 279 | 279 | 279 | 279 | 279 | 279 | 279 | 279 | 279 | 279 | 279 | 279 | 279 | 279 | 279 | 279 | 279 | 279 | 279 | 279 | 279 | 279 | 279 | 279 |
| FIDEL | Pearson Correlation | -.140* | -.157** | -.145* | -.149* | -.106 | -.164** | -.088 | -.140* | -.133* | -.114 | -.096 | -.092 | -.104 | -.062 | .027 | .196** | .027 | .101 | .235** | .458** | 1 | .466** | .352** | -.220** |
| | Sig. (2-tailed) | .019 | .009 | .015 | .013 | .077 | .006 | .144 | .019 | .026 | .058 | .109 | .125 | .083 | .302 | .649 | .001 | .649 | .091 | .000 | .000 | . | .000 | .000 | .000 |
| | N | 279 | 279 | 279 | 279 | 279 | 279 | 279 | 279 | 279 | 279 | 279 | 279 | 279 | 279 | 279 | 279 | 279 | 279 | 279 | 279 | 279 | 279 | 279 | 279 |
| MARCHIONINI | Pearson Correlation | -.143* | -.161** | -.149* | -.152* | -.108 | -.167** | -.090 | -.143* | -.136* | -.150* | -.098 | -.049 | -.024 | -.106 | -.029 | .111 | .023 | .014 | .263** | .252** | .466** | 1 | .385** | .163** |
| | Sig. (2-tailed) | .017 | .007 | .013 | .011 | .070 | .005 | .135 | .017 | .023 | .012 | .101 | .419 | .690 | .076 | .635 | .065 | .696 | .822 | .000 | .000 | .000 | . | .000 | .007 |
| | N | 279 | 279 | 279 | 279 | 279 | 279 | 279 | 279 | 279 | 279 | 279 | 279 | 279 | 279 | 279 | 279 | 279 | 279 | 279 | 279 | 279 | 279 | 279 | 279 |
| KUHLTHAU | Pearson Correlation | -.150* | -.168** | -.155** | -.159** | -.113 | -.175** | -.094 | -.150* | -.142* | -.126* | -.103 | -.054 | -.071 | -.071 | .016 | .024 | -.084 | -.074 | .392** | .142* | .352** | .385** | 1 | .580** |
| | Sig. (2-tailed) | .012 | .005 | .009 | .008 | .059 | .003 | .118 | .012 | .017 | .036 | .087 | .366 | .234 | .234 | .788 | .693 | .161 | .217 | .000 | .017 | .000 | .000 | . | .000 |
| | N | 279 | 279 | 279 | 279 | 279 | 279 | 279 | 279 | 279 | 279 | 279 | 279 | 279 | 279 | 279 | 279 | 279 | 279 | 279 | 279 | 279 | 279 | 279 | 279 |
| DERVIN | Pearson Correlation | -.130* | -.145* | -.135* | -.138* | -.098 | -.152* | -.081 | -.130* | -.123* | -.066 | -.089 | -.036 | -.052 | -.052 | .040 | .059 | -.073 | -.054 | .308** | .131* | -.220** | .163** | .580** | 1 |
| | Sig. (2-tailed) | .030 | .015 | .025 | .021 | .101 | .011 | .176 | .030 | .039 | .273 | .138 | .553 | .391 | .391 | .504 | .328 | .224 | .369 | .000 | .028 | .000 | .007 | .000 | . |
| | N | 279 | 279 | 279 | 279 | 279 | 279 | 279 | 279 | 279 | 279 | 279 | 279 | 279 | 279 | 279 | 279 | 279 | 279 | 279 | 279 | 279 | 279 | 279 | 279 |

**. Correlation is significant at the 0.01 level (2-tailed).
*. Correlation is significant at the 0.05 level (2-tailed).

Figure 8 shows the results of inputting the asymmetrical matrix into PROXSCAL for the MDS. The visualization suggests that the information retrieval researchers are more organized along a single (almost horizontal) axis than the scientometricians along a vertical one. Factor analysis of the matrix confirms this observation and makes it possible to inform this picture with a quantitative interpretation.

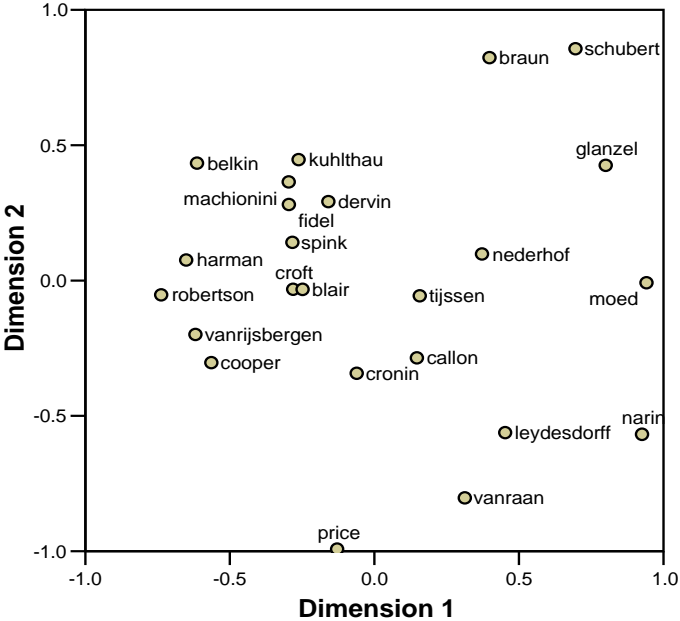

**Figure 8**:

PROXSCAL MDS on the basis of the asymmetrical matrix (normalized raw stress = 0.044)

Choosing four factors enables us to understand the relations between the two groups of authors and the fine structures within each of these groups (Table 3). The first two factors exhibit factor loadings exclusively for information retrieval researchers. These two factors explain 26.8% of the common variance in the matrix, as against 14.2% for the two factors with high loadings for the scientometric authors. This means that the information retrieval researchers are co-cited much more consistently than the scientometric authors: their co-citation patterns are more highly correlated than those of the scientometricians. The subdivisions between factors 1 and 2 and factors 3 and 4, respectively, are of a different nature. Braun, Schubert, and Glänzel are a separate group; they are mainly co-cited because of their (until recently) common address in Budapest and their many coauthored articles. The position of Cronin is special and correlates highly with that of Derek de Solla Price as a cited author. His and Price's citation patterns do not correlate specifically with any of the four factors.

**Rotated Component Matrix(a)**

|  | Component | | | |
|---|---|---|---|---|
|  | 1 | 2 | 3 | 4 |
| VANRIJSBERGEN | .679 | -.160 | -.191 |  |
| HARMAN | .652 |  | -.145 |  |
| ROBERTSON | .648 | .207 |  |  |
| CROFT | .559 | .153 | .139 |  |
| COOPER | .546 | -.168 | -.260 | -.137 |
| BLAIR | .522 |  |  |  |
| MOED | -.270 | -.229 | .216 | .231 |
| FIDEL |  | .717 |  |  |
| KUHLTHAU | -.189 | .710 | -.238 |  |
| MARCHIONINI |  | .648 | -.110 |  |
| SPINK | .267 | .568 | .164 |  |
| DERVIN | -.159 | .566 | -.236 | -.124 |
| BELKIN | .218 | .564 | -.165 |  |
| TIJSSEN |  |  | .695 |  |
| VANRAAN |  | -.123 | .615 |  |
| CALLON | -.105 |  | .478 | -.352 |
| NEDERHOF | -.111 |  | .426 | .278 |
| LEYDESDORFF | -.201 | -.179 | .426 | -.416 |
| NARIN | -.271 | -.238 | .387 |  |
| SCHUBERT | -.324 | -.280 | -.144 | .607 |
| BRAUN | -.268 | -.231 |  | .594 |
| CRONIN | -.214 | -.147 |  | -.525 |
| GLANZEL | -.265 | -.223 | .176 | .446 |
| PRICE | -.313 | -.249 |  | -.409 |

Extraction Method: Principal Component Analysis.  Rotation Method: Varimax with Kaiser Normalization.
 a  Rotation converged in 7 iterations.

**Table 3**: Factor analysis of the asymmetrical matrix of 24 co-cited authors (N = 279).

The factor analysis proceeds (by definition) from the Pearson correlation matrix as a first step in the computation of these statistics. If we enter the Pearson correlation matrix provided in Table 2 as a similarity matrix into PROXSCAL we obtain Figure 9.

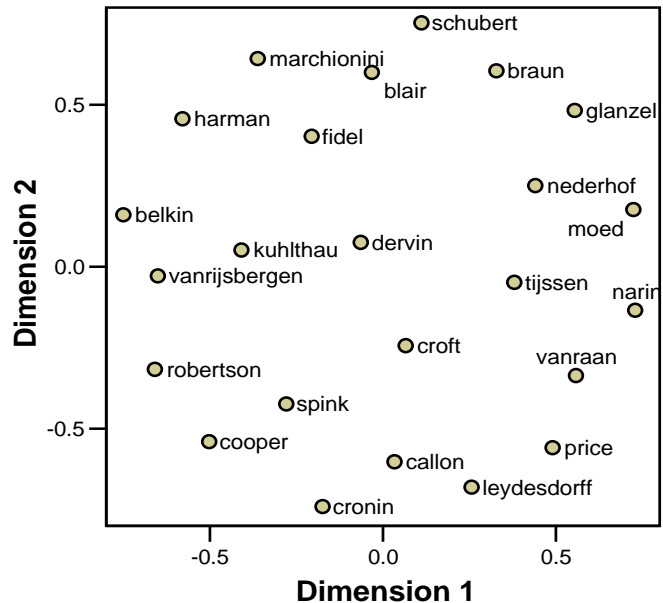

**Figure 9**:

PROXSCAL MDS on the basis of the Pearson correlation matrix provided in Table 2 (normalized raw stress = 0.148).

Although this picture enables us to make the same observation that there are two groups in this data (the information retrieval scientists on the left side and the scientometricians on the right side), the picture is less informative than the previous one and the stress has worsened considerably. The Pearson correlation matrix contains less information than the original attribute matrix because of the assumption of normality in the distribution underlying its statistics. Since our data is not normally distributed, we obtain a distorted image when we input the normalized data into MDS.[1]

By rotating the matrix, the factor analysis enables us to retrieve the underlying structure despite the assumptions made about normality in the distribution (Kim & Mueller, 1978). Furthermore, the factor analysis enables us to draw a scatter plot after optimization of the Pearson correlations with reference to the eigenvectors of the matrix. The corresponding representation in three dimensions illustrates the major division between the two groups and the fine structures within each of them.

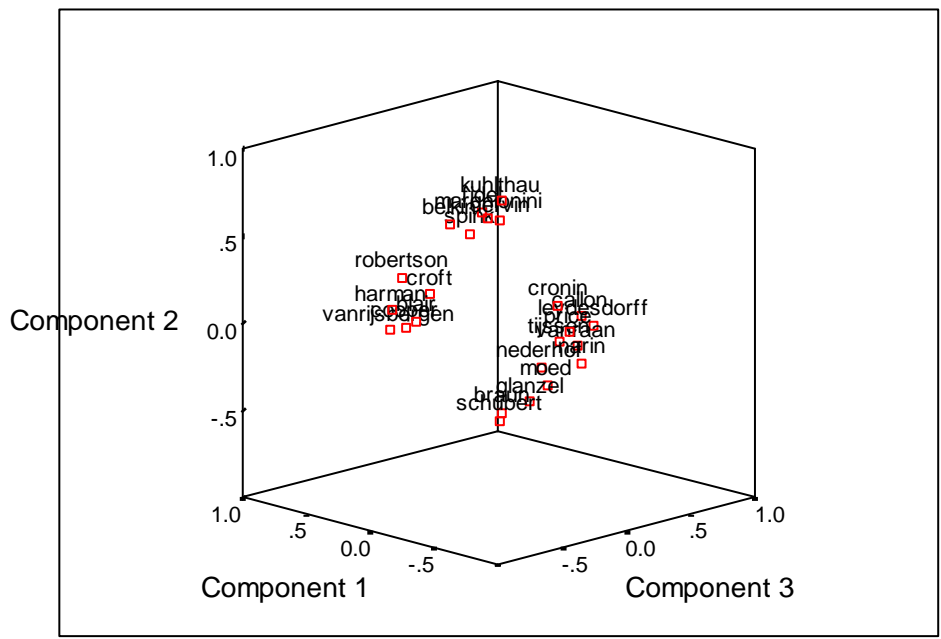

**Figure 10**:

Factor plot of the asymmetrical matrix in rotated space using three factors (Varimax Rotation; Kaiser normalization).

---

[1] The cosine is not normalized for the mean, but the normalization is otherwise strictly analogous to that of the Pearson correlation coefficient (Jones & Furnas, 1987).

*4.3    Co-citation matrix*

Similar results can be obtained by inputting the co-citation matrix directly into PROXSCAL (Figure 11). However, one has to use the co-citation data in this case as ordinal data in order to reduce the stress, that is, to improve the fit. Ahlgren *et al.* (2003, at p. 558) provide arguments to consider this data as ordinal data (Siegel & Castellan, 1988, at p. 225).[2] The split in the two groups is very clear and each group has an internal dimension corresponding to the distinctions discussed above in relation to the factor analysis.

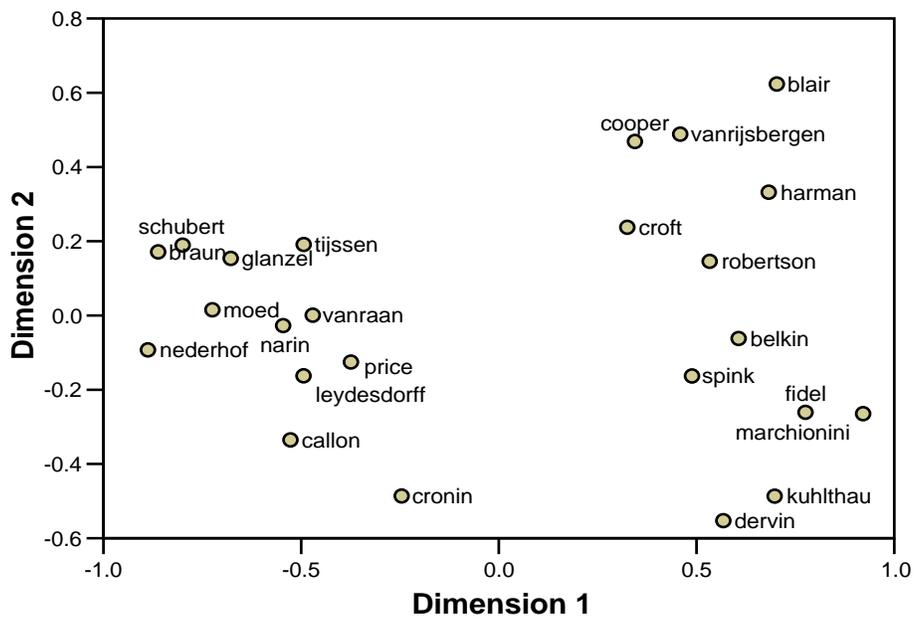

**Figure 11**. PROXSCAL on co-citation data (used as ordinal data; stress = 0.04).

When we apply this same technique to the Pearson correlation matrix based on the co-citation as input—as is common practice in ACA—we obtain Figure 12:

---

[2] Borg & Groenen (1997, at p. 170) provide arguments to "untie the observations" in the case of ordinal data. This option is available in SPSS.

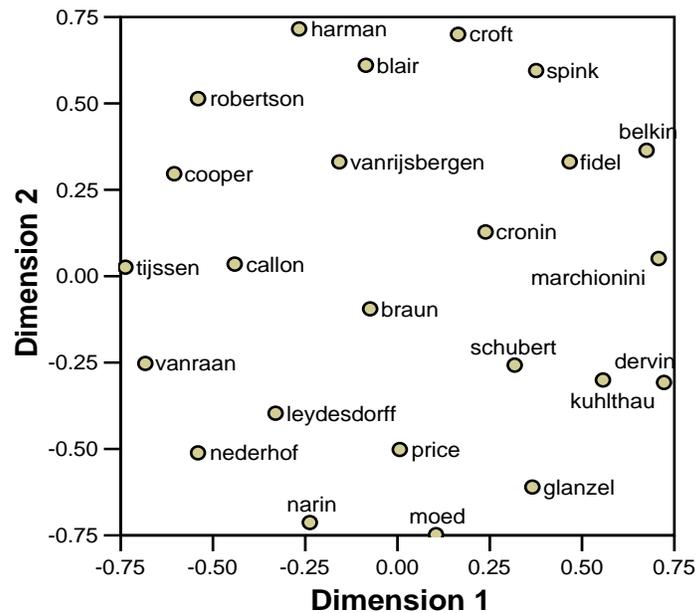

**Figure 12**:

PROXSCAL on the Pearson correlation matrix based on co-citation data (stress = 0.148).

The Pearson correlation matrix cannot be considered as ordinal data, and therefore the stress cannot be reduced by using another measurement scale. Although parts of the original structure are maintained in the solution exhibited in Figure 12, this representation of the data is less informative than the original one, and the stress is again relatively high. Part of the structure can still be recognized because of the special and very unusual nature of the data set: the two sets of authors are very distinctively different, with little overlap (Ahlgren *et al*., 2003, at p. 555).

In summary: using the Pearson correlation on a symmetrical co-occurrence matrix distorts the information contained in the co-occurrence data. If the structure in the data is robust, as is the case with this data set where the two groups of researchers are clearly separated, one may retrieve this structure nevertheless. However, this is more an exception than the norm. The example of the distances among American cities showed how applying the Pearson correlation to a symmetrical proximity matrix distorts the results even in the simple case of a geographical and therefore two-dimensional map. The advantage of using the Pearson correlation coefficient on an asymmetrical matrix becomes manifest if one proceeds to other forms of multi-variate analysis, for example, factor analysis (Bensman, 2004). However, in this case the Pearson correlation should be applied to the asymmetrical matrix, not the symmetrical co-occurrence matrix. For the representation of the symmetrical proximity matrix using MDS, one had better stay with the original (e.g., co-occurrence) matrix as input to the analysis.

*4.4. Social network analysis*

More recently, new visualization techniques have been developed in social network analysis which are based on graph theory (Scott, 1991; Wasserman & Faust, 1994).[3] While the asymmetrical matrix which we used above can be considered as a typical design in the social and behavioural sciences, these new techniques do not attribute variables (links) to cases (nodes), but study only the links and then position the nodes in terms of the links. The development of the network(s) is the subject of these studies. Co-citation data can be considered as such linkage data *among* texts, while cited references are variables attributed *to* texts. Both in terms of methodology and theoretical assumptions, however, the tradition of social network analysis and the analysis of co-occurrence data in the information sciences is somewhat different.

---

[3] For an overview of software for social network analysis see at http://www.insna.org/INSNA/soft_inf.html

We have argued above that the asymmetrical matrix contains more information than the symmetrical co-occurrence matrix. The latter can mathematically be generated from the former by multiplying the matrix with its transposed (Engelsman & Van Raan, 1991). The co-occurrence matrix can thus be derived from the original (asymmetrical) matrix and consequently contains less information (Leydesdorff, 1989). Alternatively, one can generate a dissimilarity matrix by using the Euclidean distance measure, or a similarity matrix by using the Pearson correlation or the cosine. However, network analysts are more interested in the communication structures among agents and therefore focus on links as the units of analysis. The number of links is then defined as the product of the number of occurrences among two sets. For example, if an author is cited twice in one (set of) papers and three times in another, the number of "affiliations"—as this measure is called in social network analysis—is six, while the number of co-occurrences remains only two.

This different definition does not make a difference for the mapping because the visualization algorithms in Pajek—a program which has more or less become the standard for network visualization—reduces all values first to binary values (ones and zeros) and only thereafter enables the user to visualize the values by using variable line sizes.[4] Figure 13 shows a visualization of our co-occurrence matrix using Pajek and the spring-based algorithm of Kamada & Kawai (1989). This algorithm reduces the stress in the representation in terms of seeking to minimize the energy content of the spring system. It can be considered as equivalent to non-metric multidimensional scaling.

---

[4] Pajek is freely available for academic use at http://vlado.fmf.uni-lj.si/pub/networks/pajek/ .

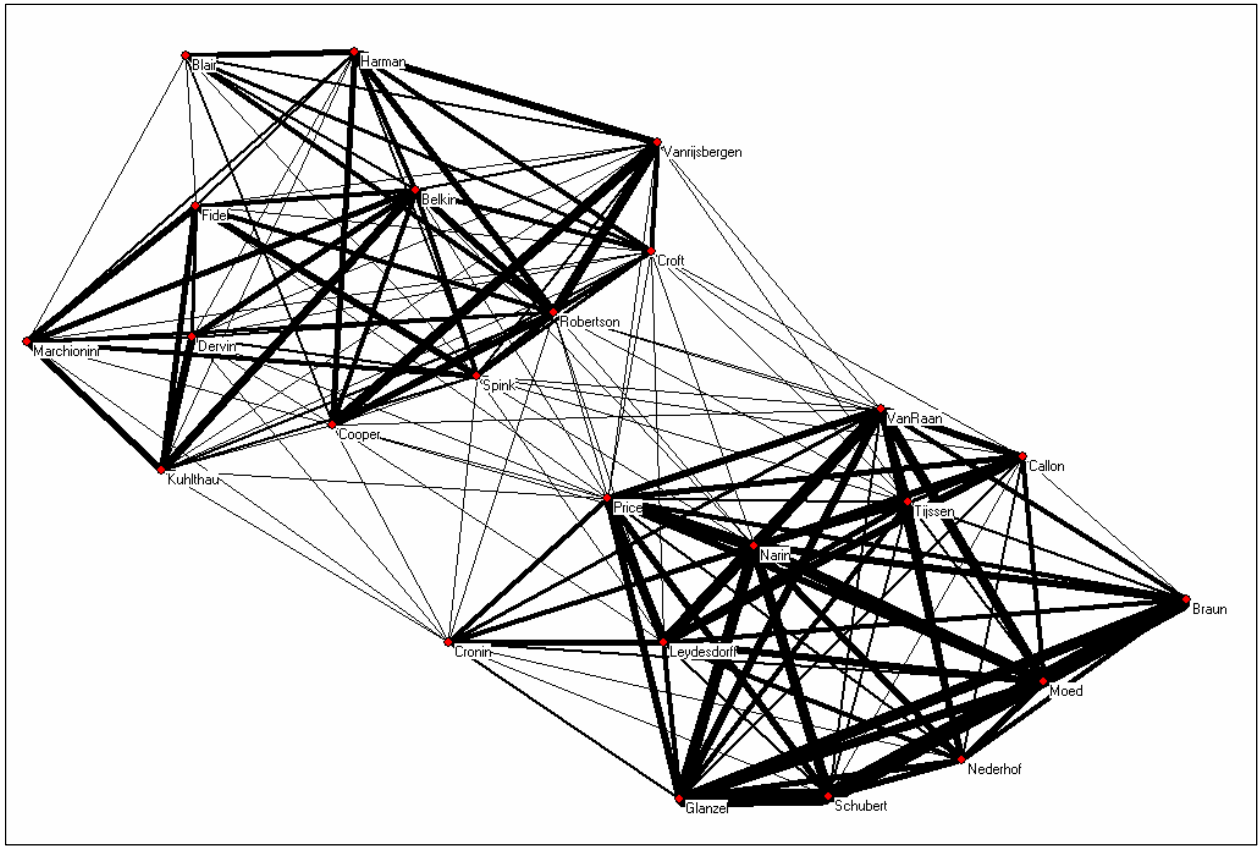

**Figure 13:** Co-occurrence matrix using Pajek for the visualization

There are several advantages of visualizing and analyzing co-occurrence data with tools from social network analysis. First, there has recently been an increasing effort to elaborate algorithms in network analysis more generally under the pressure to understand the operations of the Internet, but also of other networks in biological and physical systems (Da F. Costa *et al.*, 2005). Social network analysts can profit from these developments which are theoretically informed by graph theory. Similarly, there has been an explosion of visualization techniques.

Using PROXSCAL may be more appropriate for visualizing co-citation data than using Pajek, because PROXSCAL can take the measurement scale into account. However, Pajek allows the user to indicate the strength of the relation, as noted, in terms of the thickness of the lines. PROXSCAL and other MDS programs require users to draw the relevant lines and groupings themselves. In the above example, Figure 13 shows a mapping result similar to that of Figure 10 and Figure 11 because the data set under study is unique in that there are two distinctive sets of authors; reducing the co-citation matrix to a binary one did not affect the result significantly.

Our central message throughout this study has been that one should realize that network data are different from attributes as data. From a network perspective, for example, one may wish to focus on how the network develops structurally over time. Which functions are carried by whom at which moments of time? The scientometrician, however, is often interested in specific nodes (authors) and how they develop over time, while the network analyst may discuss this in terms of structural properties like "functional equivalents" and "structural holes" (Burt, 1982, 1995). The two traditions can be considered as different and potentially complementary perspectives on the underlying matrix. We have argued above that when the underlying matrix is available, a deeper insight into the data can be obtained by analyzing the asymmetrical matrix, but we will now turn to a case where one has no direct access to this underlying layer of nodes in the network.

## 5. The Extension of ACA to Internet Research

In the Web environment, the approach of retrieving original citation data (as shown in Figure 2) and then using Pearson's *r* to construct a similarity matrix (as shown in Figure 3) is often not feasible. The size of the Web page set is almost always too huge to be handled by individual researchers. If we are to study Web co-links, then the matrix in the form of Figure 2 is not easily available, as one would also need data on outlinks (links going out from a page). None of the existing search engines, for example, provide outlink search capability. However, several search engines such as Yahoo! and Google can search for inlinks (links going into or pointing to a site). Yahoo! also has a co-link search function which allows for data collection in the form of Figure 1. Co-link studies (e.g., Vaughan & You, 2005) parallel to those of co-citation analysis have been carried out and have found the Web to be a very useful source of data.

In the example below, we extend author co-citation analysis to the Web environment by using the search engine at http://scholar.google.com/. We searched for co-occurrences of the 24 above mentioned authors on the Web using the query strategy of the first initial and last name as recommended by Google Scholar at http://scholar.google.com/advanced_scholar_search. All searches were performed on 27 November 2004.

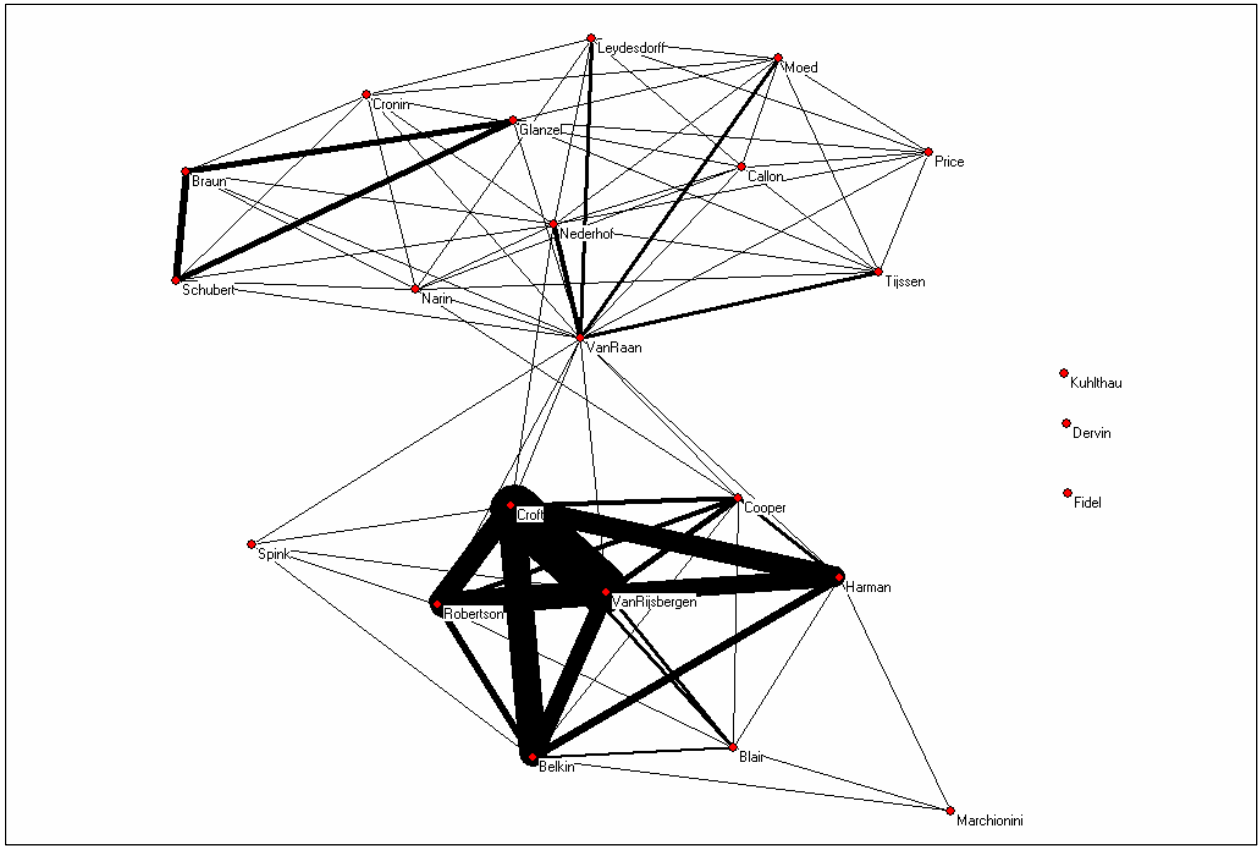

**Figure 14:**

Co-occurrence matrix using Pajek for the visualization.

Although the two groups are again very visible in this representation, Van Raan obtains the position of a hub relating the two sub-networks. Some of the information retrieval scientists do not have visibility at the Web, but some of the others are more tightly connected than the scientometricians. Within the scientometrics group, we can see Van Raan drawing mainly on a Dutch group, while the "Hungarian" group also exhibits a relatively strong relatedness.

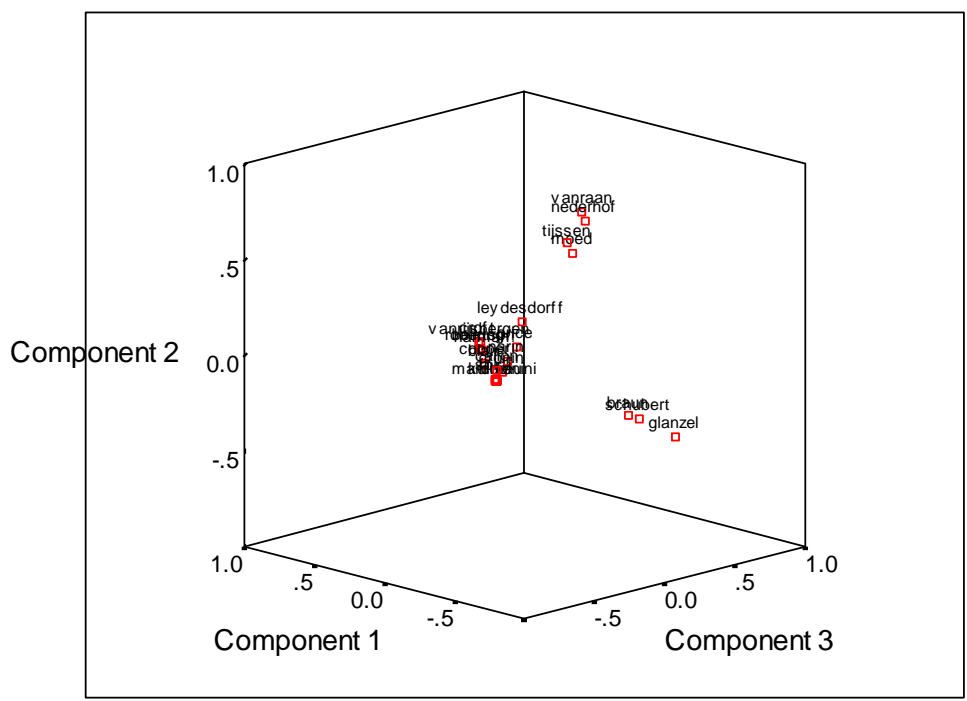

**Figure 15:**

Three-factor plot of co-occurrence file based on Google Scholar searches

These interpretations can be further informed by the factor analysis. Figure 15 illustrates the results. A first factor (explaining only 11.25% of the variance in this matrix) is again spearheaded by Croft and Van Rijsbergen. This set, however, includes also a subset of the scientometricians. The second factor (8.37%) can be considered as a "Leiden"-factor, while the third factor (6.47%) can be distinguished as the group of scientometricians with a (former) Budapest address. This pattern is somewhat different from the pattern shown in Figure 10, which was based on the ISI citation data, because the institutional component is enhanced in Figure 15.

The similarities and differences between Figure 10 and Figure 15 (that is, ISI data vs. Web data) parallel the findings from earlier studies of ISI citations versus Web citation analysis. In both library and information science (Vaughan & Shaw, 2003) and other science disciplines (Vaughan & Shaw, in press), ISI citation counts correlate with Web citation counts, but only about 30% to 40% of Web citations represent intellectual impact. The institutional and national components are thus enhanced in Figures 14 and 15 when compared with Figures 8 and 10, respectively.

A word of caution is needed, however. The stability of Web citation data is debatable (Vaughan & Shaw, forthcoming; Wouters *et al*., 2004). Furthermore, Web data are more vulnerable to manipulation than the highly codified set of the ISI (Garfield, 1979). Decades of research on ISI citation have contributed to our understanding of it, while only very limited research is available on Web citation. Although there is a rapid development of Webometrics in recent years (Thelwall, Vaughan, & Björneborn, 2005), more research in this area is needed. The extension of the co-occurrence matrix to the Web environment as discussed in this paper can be considered as an effort in this direction.

## 6. Conclusions and discussion

Co-occurrence matrices, such as co-citation, co-word, and co-link matrices, have been used widely in information science research. However, confusion and controversy persists concerning the proper statistical analysis to be applied. One root problem is in the understanding of the nature of various types of matrices. This paper has discussed the differences between the symmetrical co-citation matrix and the asymmetrical citation matrix as well as the appropriate statistical techniques that can be applied to these matrices. It concludes that the Pearson correlation coefficient should not be applied to a symmetrical co-citation matrix, but can be applied to the asymmetrical citation matrix in order to derive the proximity matrix which is needed for analysis such as multidimensional scaling. The paper also made a clear distinction between similarity and dissimilarity matrices, and we showed how they should be defined when using statistical software such as SPSS. Examples were used to support our analytical arguments.

Let us add a further reflection: although there is a standard measure of geographical distance, there is no correct way of measuring 'intellectual structure.' Both co-authorship and co-citation data are themselves only heuristics for indicating this abstract construct. Our argument, however, was not at the level of the quality of the data, that is, whether one type of co-occurrence data would be more valid or reliable as indicators of intellectual structure than another (Leydesdorff, 1989). Our message is a methodological one: if the analyst can use the underlying asymmetrical data matrix, similarity or dissimilarity can only be expressed after a proper normalization (e.g., using the Pearson correlation coefficient or Salton's cosine). The co-occurrence matrix, however, is already a summary statistic of this asymmetrical matrix: it contains less information, but can be used directly for the mapping.

One referee suggested that one could for theoretical reasons still have a preference for applying a similarity measure to the co-occurrence matrix because one would then be able to compare, for example, co-authorship *profiles* rather than co-authorship counts. We believe that the argument confuses the issue of possible limitations in the data collection stage with methodological decisions in the phase of the data analysis. If one has no data other than the co-occurrence matrix available (as in the case of Internet research), the only possibility to obtain co-authorship profiles is to input this data into statistical routines like MDS or factor analysis which begin necessarily with a proximity measure. However, one should be most cautious with applying the Pearson correlation to a co-occurrence matrix, since as we have shown above (when comparing the profiles of Van Raan and Schubert), the sign of the correlation can be changed by this manipulation of the data. If available, one should therefore prefer to use the original (that is, asymmetrical) data matrix as input to the statistical analysis. Additionally, one can derive the co-occurrence matrix from this asymmetrical matrix as another type of statistics, for example, using the routine 'Affiliations' in Pajek/UCINET, but without further processing it in terms of correlation coefficients.

The study extended the application of co-occurrence matrices to the Web environment where the nature of available data and thus data collection methods are different from those of traditional databases such as those of the ISI. A set of data collected using the Google Scholar search engine was analyzed using both traditional factor analysis and the new visualization software Pajek that is based on social network analysis. The limitations of Pajek in analyzing co-occurrence matrices were pointed out. The sole purpose of this paper is to clarify issues surrounding the nature and the application of co-occurrence matrices and thus to contribute to the further development of this area of information science.